\documentclass[11pt,twoside]{article}
\usepackage{./asp2014}

\aspSuppressVolSlug
\resetcounters

\begin{document}

\title{The Curious Case of the North Star:  the continuing tension between evolution models and measurements of Polaris}
\author{Hilding~R.~Neilson,$^1$  and Haley Blinn$^1$ 
\affil{$^1$David~A.~Dunlap Department of Astronomy \& Astrophysics, University of Toronto, Toronto, ON, Canada; \email{hilding.neilson@utoronto.ca}}}

\paperauthor{Hilding~R.~Neilson}{hilding.neilson@utoronto.ca}{}{University of Toronto}{David~A.~Dunlap Department of Astronomy \& Astrophysics}{Toronto}{Ontario}{M5S 3H4}{Canada}
\paperauthor{Haley Blinn}{haley.blinn@utoronto.ca}{}{University of Toronto}{David~A.~Dunlap Department of Astronomy \& Astrophysics}{Toronto}{Ontario}{M5S 3H4}{Canada}

\begin{abstract}
Polaris is the nearest Cepheid to us and as such holds a special place in our understanding of Cepheids in general and the Leavitt Law.  In the past couple of decades, we have learned many new things about the star as a Cepheid and as the primary component of a multiple star system.  As such, we are more precisely measuring the mass, radius and evolution of Polaris.  However, as we learn more, it is becoming clear that we understand less. There is evidence that Polaris is much less massive than stellar evolution models suggest and that Polaris is a much younger star than its main sequence companion.  In this work, we review some of the recent measurements and their connections with past studies. We then present new stellar evolution models and populations synthesis calculations to compare with the new mass measurements by Evans et al. (2018).  We find that the mass discrepancy for Polaris is about 50\%.  We also find that there is a likely age discrepancy between Polaris and its companion, but that there is also a very small probability that the discrepancy is not real. 
\end{abstract}

\section{Introduction}
Polaris is a special star.  It is the North Star and has been monitored for centuries by cultures from around the world and is an important laboratory for understanding stellar astrophysics. The star is part of a multiple star system, is a classical Cepheid, and is one of the nearest and brightest stars in the sky. This makes Polaris an ideal laboratory for modelling the structure and evolution of Cepheids.  However, measurement of the properties of Polaris continue to frustrate our understanding of Cepheid variable stars.

In this proceeding, we review some of the recent measurements of the properties of Polaris and its main sequence companion.  We then describe new detailed stellar evolution models of Polaris and its companion, and how we employ high-resolution population synthesis calculations to predict the mass and age of Polaris and its companion.  This measurement is motivated by the long history of the Cepheid mass discrepancy \citep{Cox1980, Keller2008, Neilson2011} as well as an ongoing discussion of what is the mass and evolutionary state of Polaris itself \citep{Evans2008, Turner2009, Neilson2012a, Turner2013, Neilson2014a, Bond2018, Anderson2018, Evans2018}. This study is especially motivated by new parallax measurements of Polaris~B from {\it Gaia} DR2 and revised mass measurements of the Polaris system \citep{Evans2018}.



Polaris is an important laboratory from understanding Cepheid and stellar evolution because it is part of an astrometric binary.  Polaris (or Polaris Aa) is the primary component with a F-type main sequence star Polaris Ab and a more distant companion that is also an F-type star Polaris B.  \cite{Roemer1965} provided one of the first measurements of the binary system from radial velocity observations noting the the orbit is about 30~years.  \cite{Evans2008} added Hubble Space Telescope observations that helped refine the mass measurements such that Polaris would be from about $3$-$7~M_\odot$ assuming a distance of about 130~pc from Hipparcos \citep{vanLeeuwen2007}.

That mass range did not offer strong constraints on stellar evolution models of Polaris.  Using effective temperature measurements \citep{Usenko2005} and interferometric measurements of the angular diameter \citep{Merand2006} in concert with distance measurements one could easily fit measure a mass of Polaris to be consistent with about 5~$M_\odot$ \citep{Neilson2012a}.  However, there were questions about the accuracy Hipparcos parallax and \cite{Turner2013} suggested that Polaris is actually at a distance of $99\pm2~$pc based on the argument that Polaris is a member of sparse cluster. \cite{Neilson2014a} showed that this distance and mass observation was not consistent with stellar evolution models and that models required Polaris be at least 115~pc away. 

This tension grew with independent HST parallax measurements of Polaris B \citep{Bond2018} suggesting a distance of $\approx 158$~pc.  That result created new challenges for understanding the star's evolution, requiring a larger mass for Polaris A and leading to a significant age discrepancy between Polaris A and Polaris B (and likely Polaris Ab as well).  \cite{Anderson2018} noted that Polaris A could be as massive as 7~$M_\odot$. However, this debate has become somewhat moot in light of results from Gaia DR2, where the new measured parallax of Polaris B is in agreement with the Hipparcos parallax and not the measurements by \cite{Turner2013} or \cite{Bond2018}. At the same time, \cite{Evans2018} presented results from the ongoing campaign to track the astrometric orbit in greater detail and found that the mass of Polaris A is $M = 3.45\pm0.75~M_\odot$.  This result challenges stellar evolution predictions and amplifies the apparent age discrepancy.

Polaris is also a very atypical classical Cepheid. It has a pulsation period of about 3.9 days, making it one of the shortest-period Cepheids in the Milky Way Galaxy. The star also pulsates with a very small amplitude of light and velocity \citep{Anderson2019} with a very sinusoidal variation. But, the properties that make Polaris  atypical  are related to long time-domain measurements of both its period and amplitude and attempts to understand its pulsation mode.  \cite{Neilson2012a} presented more than a century of timing measurements and found a rapid rate of period change, consistent with results from \cite{Turner2006}.  This rate could be consistent with a Cepheid on the first crossing of the Instability Strip, depending on the properties of Polaris, i.e., mass and distance \citep{Fadeyev2015, Anderson2016}.  Based on the Hipparcos distance, \cite{Neilson2012a} found that Polaris must be evolving on the third crossing and that Polaris must be undergoing enhanced mass loss.  While the rate of period change is well-established, its interpretation is not.


Polaris continues to challenge our understanding of stellar physics, whether it is through the lens of multiplicity and/or through the lens of pulsation. One of the most significant issues has been the distance to Polaris.  Understanding its distance directly impacts measurements of its mass, radius and of which crossing of the Cepheid instability strip.  The recent Gaia DR2 parallax may have solved that issue, But, this leads to two critical issues:
\begin{enumerate}
\item \cite{Evans2018} and Guzik et al.~(this proceeding) note that there is an ongoing difference between masses predicted from stellar evolution calculations and from analysis of the astrometric binary; and
\item \cite{Bond2018} and \cite{Evans2018} find a large age difference between Polaris Aa and Polaris Ab.
\end{enumerate}
We test these discrepancies using detailed stellar evolution models.
 
\section{New stellar evolution models}
We compute new stellar evolution models using the Bonn Binary Evolution Code \citep{Yoon2005}.  These models have been used for studying Cepheids and Cepheid physics previously and the details of the code can be found in previous works \citep[e.g.,][]{Neilson2011, Neilson2012a, Neilson2012b, Miller2018}.  We compute models for Polaris Aa for masses $M = 3$ to $7.1~M_\odot$ in steps of $\Delta M = 0.1~M_\odot$ that is consistent with the astrometric mass and stellar evolution calculations. We also compute each grid for assumptions of zero and moderate convective core overshooting and for initial rotation rates of 0, 25, 50, and 100~$km~s^{-1}$. We also compute stellar evolution models for Polaris Ab for a mass range from $M=1$ to $2.5~M_\odot$. We plot a sample of stellar evolution models in Fig.~\ref{f1}.

Using these models we compute population synthesis models to possible fits to the effective temperatures and luminosities for Polaris Aa and Polaris Ab.  We use the effective temperature from \cite{Usenko2005} and the luminosity from \cite{Evans2018}.  We have less information about Polaris Ab where we assume a generic effective temperature and luminosity given the color, brightness and bolometric correction.  For Polaris Aa: $\log(L/L_\odot) = 3.38\pm 0.03$ and $T_{\rm{eff}} = 6039\pm54~$K. For Polaris Ab: $T_{\rm{eff}} = 6900\pm150~$K and a luminosity $\log(L/L_\odot) = 0.74\pm0.1$.  It is unclear whether the luminosity is actually subsolar and needs to be verified.  But, if the star's luminosity is less than solar then it might be very young and offer a resolution to the apparent age discrepancy. 

\articlefiguretwo{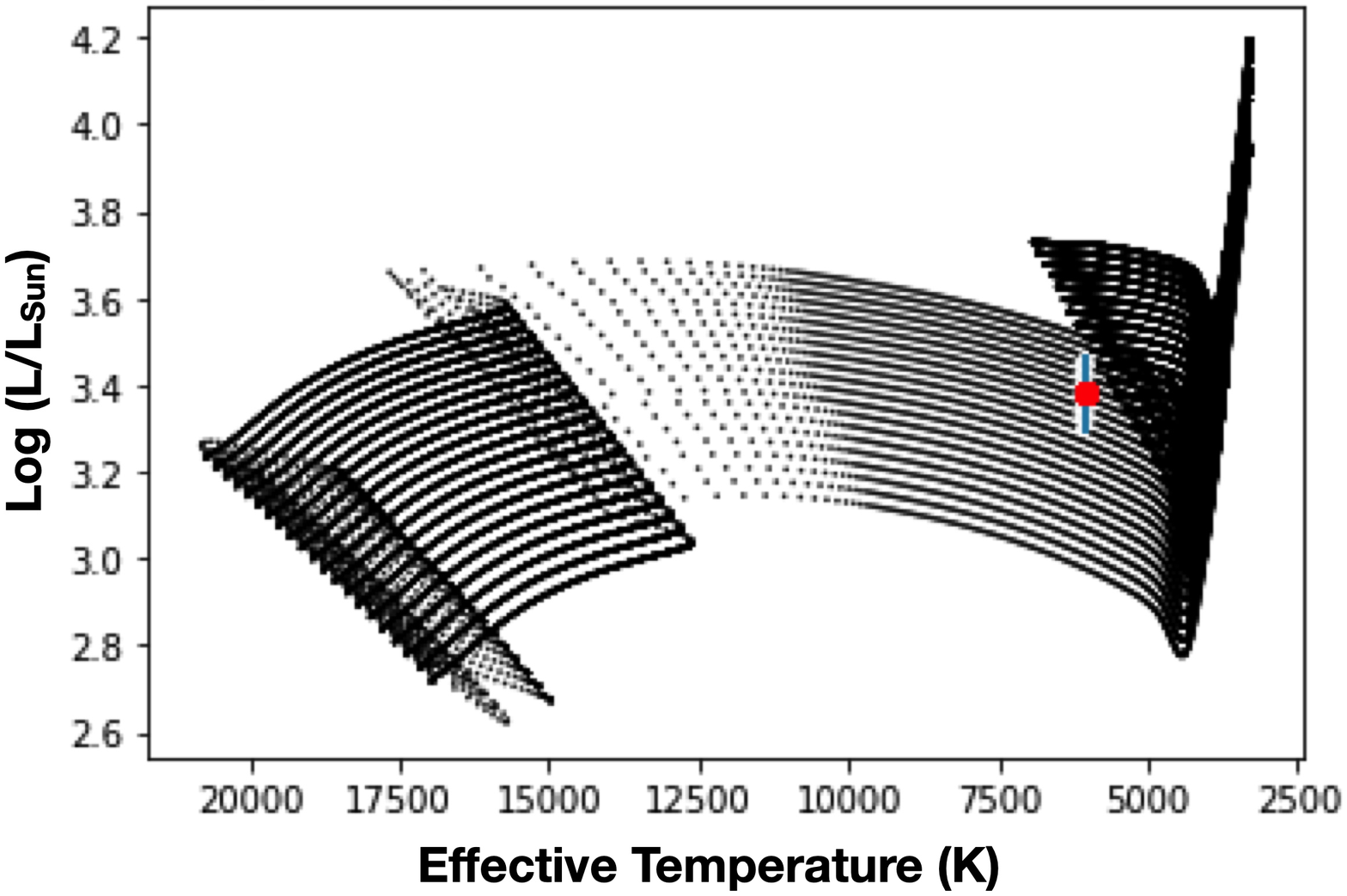}{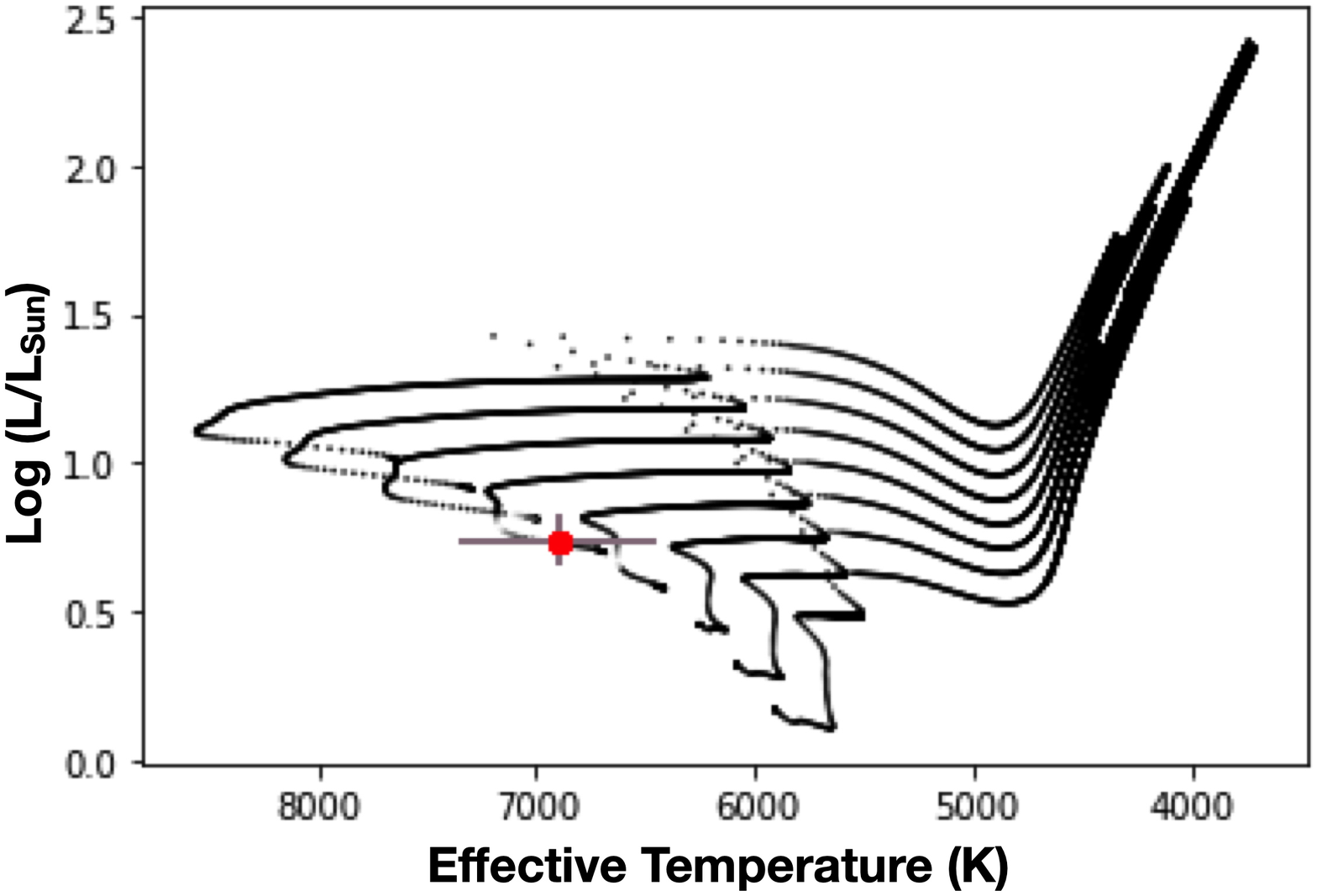}{f1}{Example of stellar evolution models for (Left) Polaris Aa and (Right) Polaris Ab for an initial rotation rate of 25~km~s$^{-1}$ and moderate convective core overshooting.}

\section{Preliminary results}
Based on our new models, we find that the mass discrepancy for Polaris cannot be resolved by testing moderate convective core overshooting and rotation. For our models we find that without convective core overshooting and zero rotation the best-fit mass of Polaris Aa ranges from $M = 5.9$ to $6.8~M_\odot$ with corresponding ages ranging from 45 to 67 Myr.  For Polaris Ab, under the same assumptions, the best-fit mass is 1.4 to $1.5~M_\odot$ with an age greater than 0.5 Gyr.  This suggests that the mass discrepancy is about 50\% when considered in the same context as \cite{Cox1980} and \cite{Keller2008}.

When we compute synthesis models for rotating stars with moderate convective core overshooting, we find similar results.  This is largely because \emph{only} one stellar evolution track evolved far enough along the blue loop to match the effective temperature and luminosity of Polaris Aa.  That was a 6~$M_\odot$ model with zero initial rotation.  All other fits were models evolving along the first crossing with masses from 6.1 to 6.8~$M_\odot$.  If the blue loops were wider, so that all models could match Polaris, then the mass would still be greater than 5~$M_\odot$.  Perhaps if Polaris were rotating along the blue loop with critical rotation then it might be possible to fit an evolution mass closer to $4~M_\odot$  \citep{Anderson2016}, but this is an extreme scenario.

For the companion, the most likely mass of the star is $1.5~M_\odot$ and an age of about 1.5~Gyr. However, for the rotating models, we find a very small probability ($\approx 1\%$) that the companion could be models by a 1.7~$M_\odot$ track with an age of about 10~Myr.  This would imply that there is a small probability that the age discrepancy is not real, but this needs to be explored in more detail.

\section{Outlook}
These results are curious and preliminary.  We find evidence that the age discrepancy suggested by \cite{Bond2018} and \cite{Evans2018} may be resolved by considering the distribution of models that could fit the properties of the companion. However, we also find that the mass discrepancy for Polaris Aa is about 50\%. 

There are a number of possible resolutions to these main issues.  The first resolution is to ask if the distance to Polaris is still unclear.  It is unlikely the Gaia parallax is incorrect, but it is anchored to Polaris B, that is one star in a multiple star system. Could the effective temperature and luminosity of Polaris be incorrect?  The effective temperature was measured from spectra \citep{Usenko2005,Usenko2018} and it is not clear if that measurement is easily reproducible.  If the effective temperature of Polaris is slightly cooler, then it is easier for blue loops to match the observations, hence suggesting a smaller stellar evolution mass.

On the other hand, perhaps our models are just wrong.  \cite{Neilson2015} resolved the age discrepancy of a Large Magellanic Cloud Cepheid binary where the Cepheid appeared much younger than the companion by modeling the Cepheid as the post-main sequence evolution of a stellar merger product.  A stellar merger acts to rejuvenate a star and make it appear younger.  Analogously, the age discrepancy might be resolved if Polaris is a merger remnant.   Given that Polaris Ab is probably 1 - 2~Gyr and Polaris Aa about 50 - 70~Myr then Polaris Aa would need to be a merger of a star with a main sequence lifetime of about 1 - 2~Gyr with another star of lesser or equal mass, hence at best a merger of a 2~$M_\odot$ star and a 1.3 to 2~$M_\odot$ companion.  This would be an unlikely scenario, but not impossible.

As a result of this work, it is challenging to draw significant conclusions beyond the fact that Polaris continues to be an enduring mystery and the more we measure the less we seem to understand. Perhaps the new decade will bring new insights that will resolve these challenges.

\acknowledgements This work was conducted at the University of Toronto that is on the traditional territory  of the Huron-Wendat, the Seneca, and most recently, the Mississaugas of the Credit River.  This work was presented on the traditional territories of Pueblo, Apache, and Dine Navajo peoples.

\bibliography{Neilson}

\end{document}